\documentstyle[aps,preprint]{revtex} 
\tightenlines                                              
\begin{document}
\draft

\title{Barrier and internal wave contributions to the quantum probability
density and flux in light heavy-ion elastic scattering}                
                                                                           
\author {F. Brau \thanks{Chercheur I.I.S.N.} and F. Michel \thanks{E-mail:
michel@umh.ac.be}}
\address{Facult\'{e} des Sciences, 
Universit\'{e} de Mons-Hainaut, B-7000 Mons, Belgium}   
                                                                          
\author {G. Reidemeister}                                                                                            
\address {Physique Nucl\'{e}aire Th\'{e}orique et Physique 
Math\'{e}matique, \\  
Universit\'{e} Libre de Bruxelles, CP229, B-1050 Bruxelles, 
Belgium}

\maketitle

\begin{abstract}

\par We investigate the properties of the optical model wave function for
light heavy-ion systems where absorption is incomplete, such as $\alpha +
^{40}$Ca and
$\alpha + ^{16}$O around 30 MeV incident energy. Strong focusing
effects are predicted to occur well inside the nucleus, where the probability 
density can reach values much higher than that
of the incident wave. This focusing is shown to be
correlated with the presence at back angles of a strong enhancement in the
elastic cross section, the so-called ALAS (anomalous large angle scattering)
phenomenon; this is substantiated by calculations of the quantum probability
flux and of classical trajectories. To clarify this mechanism, we 
decompose the scattering wave function and the associated probability flux
into their barrier and internal wave contributions within a fully
quantal calculation. Finally, a calculation of the divergence of the quantum
flux shows that when absorption is incomplete, the focal region gives a
sizeable contribution to nonelastic processes.
                   
\end{abstract}


\section{Introduction}                                                        
\label{Intro}

A better understanding of nucleus-nucleus
collision dynamics has been achieved in the last few
years by investigating light ion and light heavy-ion
systems whose scattering is not dominated by strong absorption \cite{Br97}.
When strong absorption dominates the scattering ---
the most common situation --- the scattering is known
to be sensitive to the interaction potential in the extreme
surface region only, around the so-called strong absorption
radius \cite{Sa83}. In contrast, systems that display
incomplete absorption have been found to carry information
on conditions prevailing at much smaller distances. This
information is contained in the large-angle region \cite{Br85}, as 
``anomalous large angle scattering'' (ALAS)
features at low energy \cite{De78} and clear rainbow
scattering signatures when energy increases \cite{Go72}. 
Very small absorption is not a prerequisite to the
occurence of these phenomena; indeed even in exceptional
cases such as $\alpha$-particle scattering from $^{16}$O or
$^{40}$Ca, the absolute value of the low-$\ell$ S-matrix
elements above 30 MeV incident energy is of the order of 10
percent \cite{Al82,Mi83}.

\par Although the optical model provides a satisfactory
account of many experimental data, including those displaying ALAS,
 one is often left with a
``black box'' description, where the link between the model
parameters and the calculated cross sections is rather
obscure. In the early days of the optical model,
calculations of the full scattering wave function $\psi
(\bbox {r})$ and the associated quantum flux $\bbox{j} (\bbox{r})$ 
were carried out by Mc Carthy and by Amos
\cite{Mc59A,Mc59B,Mc62,Am66,Mc6870}, in order to investigate the
scattering properties of the potential. Among other
results, these calculations revealed the importance of
focusing effects in systems such as nucleon-nucleus, where
absorption is weak \cite{Mc59A,Mc62}. The importance of the focus in
building up of a
backward peak in some transfer reactions involving protons
in the entrance or exit channel was also pointed out by
Kromminga and McCarthy \cite{Kr61}. In the case of
$\alpha$-particle
scattering, where absorption is stronger, it
was found at that time \cite{Mc59B} that, even when it could be
discerned in the far side region, such a focusing effect
has a negligible influence on the scattering, since
propagation of the flux out of the nuclear
medium leads to a nearly complete extinction of this
contribution. Moreover, 
not much physical significance was attributed to the
scattering wave function inside the nuclear volume, since
the status of the optical model potential for composite
particle scattering was then still
very obscure.  Surprisingly enough,
few calculations of this type were reported subsequently in
the literature; they have however been revived now and then
in various contexts \cite{Ma88A,Ma88B,Ma88C,Da95}.

\par One obvious drawback of this type of approach is that
the scattering wave function, and derived quantities
like the quantum flux, contain contributions from all the
mechanisms that are possibly active in the scattering
system under study. Therefore many techniques have been
developed to try to understand particular features observed
in the cross sections in more familiar terms. Semiclassical
approaches (as in Ref. \cite{Br85} and references therein) have
played a key role in this respect, even for systems where
the applicability of these methods could seem problematic. 
Concepts like rainbow or glory scattering, orbiting and spiral scattering
\cite{Fo59}, nearside and
farside \cite{Fu75}, or internal wave and barrier wave contributions
\cite{Br77},
 have thus become
commonplace in the optical model literature. One is thus
led to the somewhat paradoxical situation where, although
numerically exact results can be obtained from the
formalism, the latter often contains less useful information than
approximate solutions.

\par The philosophy of many of these approaches is to
decompose the scattering amplitude into several
subamplitudes with, hopefully, simpler properties. For
example, in one of these approaches \cite{Fu75,Hu84}, the
elastic scattering amplitude is decomposed into the
so-called nearside and farside amplitudes. The cross sections
corresponding to each of these contributions are generally
smooth, and their interference explains the Fraunhofer
diffractive oscillations seen in various heavy-ion elastic
scattering angular distributions. In another approach \cite{Br77},
the scattering amplitude is decomposed into its barrier
and internal components. In contrast
to the previous method, this approach works best at low
energies, where the effective potential displays for each
active partial wave (provided the nuclear potential is deep enough) a
potential pocket separated  by a
barrier from the external region. This approach has
explained how the anomalous features (ALAS)
observed in elastic scattering for some light-ion systems
emerge from an optical model description when the real part
of the potential is deep and absorption is
particularly low \cite {Br85}; in particular, it has
definitely settled the surprising fact that, in admittedly
exceptional cases, part of the incident flux can remain in
the entrance channel after a deep excursion into the
nuclear medium.   

\par The information obtained from semiclassical approaches
can sometimes be obtained by resorting to purely quantal
methods. For example, it has been shown that the
semiclassical barrier-internal wave decomposition of the
elastic scattering amplitude by Brink and Takigawa
\cite{Br77},
initially carried out within a WKB approximation context,
could be performed by resorting to ordinary optical model
calculations \cite{Al82}. In its simplest version, the
technique consists in enhancing
artificially the absorptive potential in the inside region
of the potential, in order to make the internal wave
contribution negligibly small, which provides the barrier
wave contribution; the latter is subsequently subtracted
from the full
amplitude to calculate the internal wave component. An
advantage of this approach is to provide --- in contrast
to the semiclassical calculations whose basic ingredients
are action integrals evaluated between the active turning
points \cite{Br85,Br77} --- wave functions corresponding to the different
contributions to the scattering amplitude. 
 
\par In view of the importance of a better understanding
of the mechanisms underlying light ion and light heavy-ion
scattering, we have reinvestigated the properties of the
elastic scattering wave function and the associated quantum
flux for a few light-ion systems. Our results can be summarized as
follows. In all cases we have studied, focusing effects are indeed observed
at low energy. When absorption is incomplete, focusing can
become very strong and the probability density at the focus is found to
reach values much larger
than that of the incident wave --- in some cases even 
larger than the values reported by McCarthy \cite{Mc59A,Mc62} and by Amos
\cite{Am66} for
low energy nucleon scattering. The presence of the
focus, which is located well inside the nuclear medium at
low energy, is then found to be correlated with the
occurence at large
angles of an internal wave contribution that dominates the
scattering in the backward hemisphere and is responsible
for the ALAS phenomenon. This is clearly demonstrated by
examining the properties of the internal wave contribution
to the total wave function. This focus thus appears to be
the region of the nuclear medium from which most of the
internal wave contribution to the elastic scattering cross
section originates when absorption is incomplete.

\par As energy increases, the focus is found to
move away from the center towards the far side of the
nucleus; accordingly the
flux that is refracted at back angles decreases --- an
effect which is enhanced by the increase of absorption with
energy --- and glory scattering is progressively replaced
by rainbow scattering.

\par On the other hand, the calculation of the divergence
of the quantum flux, which indicates where absorption is
most effective, shows that the latter occurs in two
distinct regions: in the surface region, on the illuminated side
of the nucleus, and, further inside the nucleus, around the
focal point. For strongly absorbing systems, the first
mechanism is clearly dominant, in accordance with the
models
generally used in direct nuclear reactions calculations
which locate most of the coupling strength in the
surface of the target nucleus, while for systems displaying reduced
absorption, the second mechanism is significantly enhanced.

\par This paper is organized as follows. In Section
\ref{focus}, we compare probability densities calculated for several
optical model potentials, both for systems presenting reduced or strong
absorption, as well as the associated quantum probability fluxes; we also
investigate the energy dependence of the focusing properties of these
potentials. In Section \ref{decomposition}, we decompose the scattering
wave functions into barrier and internal wave components, thus
obtaining the contribution of these two components to the probability density
and the probability flux for these systems; the importance and localization
of absorption are studied by calculating the barrier and internal wave
contributions to the divergence of the quantum flux. A summary is
presented in Section \ref{summary}.

\section{Focusing properties of optical potentials}
\label{focus}

\subsection {A historical perspective}
\label{history}

\par The measurement of low-energy elastic
$\alpha$-particle scattering up to large angles has
disclosed the existence, for a few light targets like
$^{16}$O \cite{Mi83,Ab93} and $^{40}$Ca \cite{De78,Gu81}, of anomalous
features in the angular distributions: whereas in many cases the angular
distributions remain diffractive on the whole angular
range, a large rise of the cross section is observed at back angles for
these targets; around 30 MeV incident energy, this rise can
exceed the Rutherford cross section by two to three orders
of magnitude. When the energy increases, this backward rise,
termed ALAS, disappears progressively and is
replaced around 100 MeV incident energy by a rainbow behavior. 

\par It was found that these anomalous features, which were
long thought to lie outside the capabilities of an optical
model
description, can be reproduced quantitatively by
using optical potentials with an imaginary part distinctly
weaker than that used for ``normal'' systems and with a
real part described by a conveniently chosen form factor
\cite{Mi77,De78,Mi83}.
It was soon realized that the existence of a backward rise
in the cross section is due to part of the incident
wave that crosses the effective potential barrier and
reemerges after having been reflected at the innermost
turning point \cite{Br77}, and thus that, contrary to what had been
considered to be a general rule, elastic scattering of
composite particles like the $\alpha$-particle is not
necessarily governed by strong absorption. An important
consequence of this unexpected transparency is that the experimental
elastic scattering cross sections carry informations on the
interaction potential well inside the strong absorption
radius, and indeed a consistent study of the phenomenon on
a broad energy range makes possible the extraction of an
unambiguous global optical potential whose real part is definitely
deep and is well defined up to fairly small distances \cite{Gu81,Mi83}. 

\par These potentials were later shown to contain more
than a simple parametrization of the cross sections. Indeed 
the properties of the phenomenological $\alpha + ^{16}$O
global optical potential were shown \cite{Mi83,Wa87} to be compatible with
microscopic
approaches such as resonating group method (RGM), which
take into account antisymmetrization effects between
projectile and target in an exact way. In particular, the
numerous unphysical states, which are bound by the deep
phenomenological potentials below the threshold, were shown
to be close analogues of the so-called forbidden states of
the RGM, and must thus be discarded \cite{Ho91}. One is thus led to give
credibility to the wave functions associated with the deep local
potentials obtained from analyses of elastic light-ion
scattering down to small distances, the more so as the
effects of non-locality on the wave
function (the so-called Perey effect) are known to be small
for low-mass projectiles \cite{Ho91}.

\subsection {Comparison between strong absorption and
reduced absorption}
\label{absorption}

\par As a starting point, we investigate the properties of
two optical potentials \cite {Al82} describing
$\alpha$-particle scattering from targets of comparable
masses at the same incident energy, that is, $^{40}$Ca
and $^{44}$Ca at 29 MeV. The main difference between these
two systems lies
in the strength of the absorption needed for describing the
data: whereas $\alpha$ particles scattered from $^{44}$Ca
are strongly absorbed and the angular distribution
displays a diffractive behavior up to large angles, the
$\alpha + ^{40}$Ca system is characterized by an incomplete
absorption and a spectacular backward enhancement. The
angular distributions calculated with these two potentials,
which give a good description of the
experimental angular distributions over the whole angular
range, are contrasted in Fig.~\ref{sections4044}. 

\par The probability densities $\rho (\bbox{r}) = |\psi(\bbox{
r})| ^2$ 
associated with these two optical potentials are
displayed in Fig.~\ref{fo4044} and are seen to be generally
similar. In particular, the ``parabolic cup'' surrounding
the interaction region is essentially a Coulomb effect. In
the forward direction, the structures observed for the two
systems are also nearly identical outside the
interaction region. This is not so at larger angles: one
observes the apparition of ripples on the illuminated
side of the $^{40}$Ca nucleus in several preferred
directions, especially around $\theta =
180^\circ$, which (as shown later) are due to an internal wave contribution to the
scattering, whereas  the $\alpha + ^{44}$Ca
probability density is essentially flat on the illuminated side. Moreover, a strong
focusing effect is seen to be present behind the center of
the nucleus in the $^{40}$Ca case, whereas it is
barely visible in $^{44}$Ca. (Note the use of a logarithmic
scale in the figure). This focus is followed at larger
distances by a broad ridge whose importance is seen, in
contrast, to be barely affected by absorption.

\par To give a more quantitative
impression of these effects, we present in Fig.~\ref{sagittal4044} a section
of the probability densities along the axis of the incident
beam ($z$ axis). It is clearly seen that the focus is
localized well inside the interaction region, at about 2 fm from 
the center of the target nucleus. This behavior
is similar to that reported by McCarthy in his analysis of
low-energy neutron scattering \cite {Mc59A,Mc62}. Whereas the
magnitude of the
peak at the focus in the $^{44}$Ca case is lower than that
of the
incident wave, it reaches about 20 times that value in the
$^{40}$Ca case. In contrast, the broad ridge alluded to
above is seen to develop mainly outside the interaction
region. Finally, the oscillations observed on the illuminated side
of the $^{40}$Ca nucleus, which will be shown to be related to the internal
wave contribution to the scattering, are seen to be
strongly suppressed in $^{44}$Ca.

\subsection{Quantum probability flux and classical trajectories}
\label{trajec}

\par To understand better the origin of the
features seen in the density plots, we calculated the
quantum flux 

\begin{equation}
\label{flux}
\bbox {j} (\bbox {r})= \frac{\hbar}{\mu} \Im \left( \psi^\ast (\bbox {r})
\bbox{
\nabla}\psi(\bbox {r})\right)
\end{equation} 

associated with the total wave function $\psi (\bbox{r})$ for
the two cases presented above (Figs.~\ref{flux40} and~\ref{flux44}).  For
large impact
parameters the incident flux does not penetrate into the
nuclear interaction region and one observes a bunching of
the streamlines at the edge of the parabolic cup
mentioned above, which is clearly associated with the
Coulomb interaction. For smaller impact parameters, the
streamlines are progressively pulled towards the nuclear
center and the flux vectors are seen to converge to a
region located near the focus observed in the probability
density. Whereas the flux density is seen to be rapidly
damped on its way towards the focus in the $^{44}$Ca case,
and as a result the intensity at the focus remains
rather
small, in the $^{40}$Ca case this intensity is seen to
increase significantly, reaching a much higher value at the
focus. 

\par Unfortunately, it is difficult to obtain more insight
from this figure, because the flux calculated here includes
both the incident wave and the scattered wave
contributions. 

\par As was shown by McCarthy in his pioneering
calculations, the classical trajectories are useful 
for investigating qualitatively the focusing properties of
the potential,
provided the incident energy is not too low \cite{Mc59A}. The classical
trajectories associated with the real part of
the 29 MeV $^{40}$Ca potential for a few impact
parameters are shown in Fig.~\ref{trajectories}. It is seen that the
trajectories with an impact parameter less than about 6
fm converge in a precise way to a point located very near
the quantum focus. These classical trajectories are identical
to the rays which would be calculated in a geometrical
optics context using the position-dependent refractive index

\begin{equation}
\label{index}
\ n(r) = \sqrt{1-\frac{V(r)}{E_{\rm c.m.}}}
\end{equation}

where $V(r)$ denotes the real part of the optical potential.

The refractive index near the origin for the incident energy and the
potential considered here is comparable to
that of diamond for ordinary light, that is about 2.5;
this is why focusing occurs inside the refracting sphere at
low energy. When the energy increases, Eq. (\ref{index}) predicts a
decrease of the refractive index; in this simple picture
the focus is thus expected to shift away from the nuclear
center as energy increases, if one assumes that the
potential depth is energy independent (which is indeed the
case in a first approximation), a feature already
observed by McCarthy in his calculations.

\subsection{Energy dependence of the focusing properties of
the potential}
\label{energy}

\par To conclude this tour of the focusing properties of
the $\alpha$-nucleus optical potential, we examine the
energy behavior of the probability densities. We
concentrate here on another transparent system which has played a key role
\cite{Mi83,Ho91} in understanding the dynamics of the $\alpha$-nucleus 
interaction, the $\alpha + ^{16}$O system. The
parameters used are those of the global optical
potential in Ref. \cite{Mi83}. As is seen in Figs.~\ref{energydep} and
~\ref{sagittalo16}, the focus moves away from the illuminated side of the 
nucleus when energy increases, and the density at the focus decreases
steadily. These properties are easily understood if one
takes into account the fact that the real potential depth
at small distances decreases slowly with energy,  ranging 
from about 160 MeV to 120 MeV when the
incident energy increases from 30 to 150 MeV \cite {Mi83}, while
absorption increases regularly in this range. The refractive
index in Eq. (\ref{index}) thus also decreases with energy; for
$^{16}$O$(\alpha,\alpha)$ scattering, it varies from about
2.7 to 1.4 over the same energy range and the focal length
of the system increases accordingly. It is interesting to note that the
region of the potential to which the scattering is most sensitive, which was
obtained in Ref.\cite{Mi83} from a
notch test analysis, coincides with the location of the focus at low energy.

\par Above about 60 MeV, we found that the low angular
momentum
classical trajectories are still converging to a focus
inside the
target nucleus but, contrary to the example displayed in
Fig.~\ref{trajectories},  they are not deflected beyond some
critical angle that decreases with energy . Accordingly, ALAS is
progressively replaced by a rainbow behavior and
the ripples that were still clearly observed on the illuminated side
of the target at 32.2 and 49.5 MeV are seen to have
completely disappeared by 69.5 MeV as a result of the
disappearance of the internal wave contribution to the
scattering beyond this energy.

\section{Barrier-internal wave decomposition of the wave
function}   
\label{decomposition}

\subsection{The barrier and internal wave contributions to
elastic scattering}
\label{contrib}

\par In order to clarify the focusing properties of the investigated
nuclear potentials, we have
decomposed the elastic scattering wave function into two
contributions, corresponding respectively to the part of
the incident flux reflected at the barrier of the
effective potential and the part that penetrates the
nuclear interior. This decomposition makes sense for the systems studied here
at low energy, since the effective potentials have a pocket for all the
active partial waves. It must be stressed that in the original
semiclassical internal-barrier wave decomposition of Brink
and Takigawa \cite{Br77}, this decomposition is not performed on
the scattering wave function but on the scattering
amplitude $f(\theta)$, making possible the calculation of
``barrier'' and ``internal wave'' contributions
$f_B(\theta)$ and $f_I(\theta)$ to $f(\theta)$, and thus of the contributions

\begin{equation}
\label{Crossections}
\ \sigma_B(\theta)=|f_B(\theta)|^2, \;\;\;\; \sigma_I(\theta)=|f_I(\theta)|^2
\end{equation}

to the full elastic scattering cross section $\sigma(\theta)$.
More precisely, the
elastic scattering matrix $S_\ell$ is written as \cite{Br77}

\begin{equation}
\label{smatrix1}
\ S_\ell=S_{B,\ell}+S_{I,\ell}
\end{equation}

where $S_{B,\ell}$ is given by 

\begin{equation}
\label{smatrix2}
\  S_{B,\ell} = {\exp (2i \delta_1^\ell)}/{N_\ell}    
\end{equation}

and, if multiple reflections between the two inner
turning points (that is, resonances in the potential
pocket) are neglected, a condition which is met in most
cases except possibly at very low incident energy,
$S_{I,\ell}$ is given by 

\begin{equation}
\label{smatrix3}
\ S_{I,\ell} = {\exp (2i \delta_3^\ell)}/{N_\ell^2}
\end{equation}

\par In Eq. (\ref{smatrix2}), $\delta_1^\ell$ is the usual WKB phase
shift
corresponding to the external turning point and $N_\ell$
measures the penetrability of the barrier of the effective
potential $U_{\rm eff}^\ell$ for angular momentum $\ell$; in Eq.
(\ref{smatrix3}), $\delta_3^\ell$
is the WKB phase shift corresponding to the  innermost
turning point

\begin{equation}
\label{smatrix4}
\delta_3^\ell=S_{32}^\ell+S_{21}^\ell+ \delta_1^\ell
\end{equation}

where $S_{ij}^\ell$ denotes the semiclassical action
integral
for angular momentum $\ell$, evaluated between the
(complex) turning points $r_{i,\ell}$ and
$r_{j,\ell}$

\begin{equation}
\label{smatrix5}
\ S_{ij}^\ell= \int_{r_{i,\ell}}^{r_{j,\ell}} dr \left\{
\frac {2 \mu}{\hbar^2} \left[ E_{\rm c.m.}-U_{\rm eff}^\ell \right]
\right\} ^{1/2}  
\end{equation}

Finally, the full elastic scattering amplitude $f(\theta)$
is decomposed as \cite{Br77}

\begin{equation}
\label{smatrix6}
\ f(\theta) = f_B(\theta) + f_I(\theta)
\end{equation}

where the barrier wave and internal wave amplitudes, $f_B$
and $f_I$, are given in conventional notation by

\begin{eqnarray}
\label{smatrix7}
\ f_B(\theta) &  = &  f_R(\theta) + \frac{1}{2ik}\sum_\ell
(2\ell+1) \exp(2i \sigma_\ell) \left[ S_{B,\ell} -1 \right]
P_\ell (\cos \theta) \\
\ f_I(\theta) &  =  &   \frac{1}{2ik}\sum_\ell  (2\ell+1)
\exp(2i \sigma_\ell)  S_{I,\ell}  P_\ell (\cos \theta)
\end{eqnarray}

\par This decomposition, which requires the localization
for each $\ell$ value of the active turning points and the
evaluation of action integrals in the complex plane between
these turning points, seems in principle to be restricted
to analytical potentials. It was however shown in Ref. \cite{Al82}
that it can in fact be carried out in a fully quantal context,
using scattering matrix coefficients supplied by any
optical model code. The basic technique consists in
enhancing artificially the absorption at small distances to
enhance the imaginary part of $S_{32}$, in order to
damp the internal wave contribution to the scattering
amplitude and thus to provide the barrier wave contribution
$f_B(\theta)$. The internal wave amplitude $f_I$ is
obtained in a second step by subtraction of $f_B$
from the full scattering amplitude $f(\theta)$. The extra
absorption used must of course be restricted to small
distances in order to preserve both the external WKB phase
shift $\delta_1^\ell$ and the barrier penetration factor $N_\ell$.
Although, as explained in Ref. \cite{Al82}, this simple technique
leads in most cases to good agreement with the full WKB
calculation, it was found to lead sometimes to serious
discrepancies, and therefore a more elaborate quantum
mechanical scheme was devised in Ref. \cite{Al82} in order to
alleviate these
problems. In the rest of the present paper, we will 
use the simpler technique described above, since for the
cases examined here it proved quite stable and reliable.

\par An important byproduct of this technique is to
provide, beyond the $S$-matrices $S_B$ and $S_I,$ wave
functions $\psi_B$ and $\psi_I,$ defined even in the
interaction region, associated with the 
barrier wave and internal wave contributions. Use of a conveniently enhanced
absorption provides the barrier contribution $\psi_B$ to the total wave
function $\psi$, and the internal contribution $\psi_I$, which we define by

\begin{equation}
\label{psiib}
\ \psi=\psi_B+\psi_I
\end{equation}

is thus obtained in a second step by subtraction of $\psi_B$ from
$\psi$. 

\par Although fine details of the components of the wave function thus
obtained depend
somewhat on the exact prescription used for enhancing the absorption, we
checked that --- as far as the
WKB component $S$-matrices are correctly reproduced --- little uncertainty
arises in our decomposition of the wave function.

\subsection{Components of the wave function}
\label{influence}

\par We applied the technique described in the
previous subsection to the $^{40,44}$Ca$(\alpha,\alpha)$
cases
at 29 MeV. The potential parameters are still those of
Ref. \cite{Al82}; for the imaginary potential $\Delta W(r)$ needed
to enhance absorption in the internal region, use was made
of the same form factor as the perturbative potential used
in that work, that is

\begin{equation}
\label{Deltaw1}
\Delta W(r) = \Delta W_0  \exp \left[ -(r/\rho)^4 \right]
\end{equation}

As discussed in Ref. \cite{Al82}, an adequate choice of the
parameter $\rho$ guarantees that this form factor decreases
sufficiently rapidly in the barrier region, a feature that
is important to avoid unwanted modifications of the barrier
contribution. A convenient choice is 

\begin{equation}
\label{Deltaw2}
\ \rho \approx R_B/2
\end{equation}

where $R_B$ denotes the barrier radius at the grazing
angular momentum. At the same time, this form factor decreases
sufficiently smoothly so
as not to introduce additional spurious turning points in
the problem. The results of the calculation should not depend critically on these cutoff parameters;
the values used here are $\Delta W_0$  = -100 MeV, $\rho =
3.25$ fm for the $^{40}$Ca case, and $\Delta W_0$  = -50
MeV, $\rho = 3.40$ fm for the $^{44}$Ca case.

\par The barrier wave and internal wave cross sections
$\sigma_{B}(\theta)$ and $\sigma_{I}(\theta)$ corresponding
to these two systems are compared in Fig.~\ref{decompositions}, together with
the moduli of the corresponding $S$-matrix coefficients
$S_{B,\ell}$ and $S_{I,\ell}$. One sees that while the
barrier wave contributions are remarkably similar for both
systems (except for trivial size effects), the internal
wave contributions to the $S$-matrix have the same cutoff
angular momenta, but differ by about one order of
magnitude. Correspondingly, the internal wave cross sections are seen to
differ by about two orders of magnitude, but they have a
very similar pattern. The ALAS phenomenon observed in the
$^{40}$Ca case is thus seen to be entirely due to an
enhanced internal wave contribution to the scattering, as
was first established by Brink and Takigawa \cite{Br77}.

\par The very different role played by these two
contributions is beautifully illustrated by the 
probability densities $|\psi_B|^2$ and $|\psi_I|^2$
obtained for the two systems, which are displayed in Figs.~\ref{decompfo40}
and \ref{decompfo44}. Again the barrier probability
densities are seen
to be strikingly similar for both systems. We note in
passing that the broad ridge observed in the very forward
direction, which has a comparable importance in the two
systems, and which should not be confused with the focus
found inside the nucleus, is essentially a barrier
phenomenon. 

\par In contrast, while the internal probability densities
have a very similar pattern, they differ by about two
orders of magnitude. They both display a prominent peak
located behind the center of the target, which coincides
with the focus observed in the full scattering wave
function (see Fig.~\ref{fo4044}); of course the $^{44}$Ca focus is
about two orders of magnitude lower than its $^{40}$Ca counterpart.
This peak is preceded on the illuminated side by a broad bump centered
around the origin; the latter was not conspicuous in the
full wave function because the barrier contribution is
still important in this region. At larger distances, the
internal density is seen to oscillate in the backward
hemisphere; the angular positions of its maxima and minima coincide with
those of the internal wave contribution to the cross section
(Fig.~\ref{decompositions}), and
thus with those of ALAS in the $^{40}$Ca case (Fig.~\ref{sections4044}).  A
more quantitative comparison of the different components of the
wave function along the axis of the incident beam can be
found in Fig.~\ref{sagittalca}.

\par One of the merits of our decomposition of the wave
function is to display in a striking way the strong
correlation between the existence of a focus inside the
target nucleus and an internal wave contribution to
the scattering cross section. When absorption dominates the
scattering, a focus can still be discerned in the
internal density, but its contribution to the total density is
comparatively weak and its contribution to backward
scattering is negligible. In contrast, in a context of
incomplete absorption, the focus is found to play a leading
role in the building up of the ALAS phenomenon observed in
the backward angular distribution.

\subsection{Components of the quantum flux}
\label{influence2}

\par We have likewise calculated the quantum flux
corresponding to each of the wave function components for
the same two systems; these flux components, which will be denoted by $\bbox
{j}_B$  and $\bbox {j}_I$, are calculated from
Eq. (\ref{flux}) using the barrier or the internal component of the wave
function. We note that $\bbox {j}$ and $\bbox {j}_B$, which both
derive from wave functions satisfying a Schr\"{o}dinger equation with an
absorptive potential, have necessarily a negative divergence. This is not
necessarily so for $\bbox {j}_I$ since the equation for $\psi_I$, which reads

\begin{equation}
\label{Schro}
\ -\frac {\hbar^2} {2m} \bbox{\nabla}^2 \psi_I + (V+iW) \psi_I - i \Delta
W \psi_B = E \psi_I
\end{equation}

(where $V+iW$ is the original optical potential and $\Delta W$ is the extra
absorption of Eq. (\ref{Deltaw1})), is coupled to $\psi_B$.

For the $\alpha + ^{40}$Ca
system, we present in the lower part of Fig.~\ref{fluxint40} a closeup of the
internal flux
contribution in the focus region, which essentially
confirms the features observed for the total flux in the
lower part of Fig.~\ref{flux40} for that system. It is, however,
interesting to notice
that the rather peripheral current lines which bend towards
the axis and contribute to the enhancement of the total
wave
function beyond 10 fm (see Fig.~\ref{flux40}) are not present
here and that they are thus clearly associated with the barrier wave
function. On the other hand, we show in the upper part of
Fig.~\ref{fluxint40} the
internal flux contribution on a broader scale; it has been
multiplied by a factor of 1000 in order to enhance its
asymptotic behavior 
and has not been represented for distances lower than 6
fm, where it is much larger and would overflow the figure
at this scale.
The full line and dashed line circles represent the
distance where the real and the imaginary parts of the
optical potential have fallen to one tenth of their central
values. 

\par In Fig.~\ref{fluxbar40}, we finally display the barrier part of
the quantum flux for $^{40}$Ca$(\alpha,\alpha)$ scattering;
one sees in this figure how the current lines grazing the
surface of the potential survive absorption to build up
a sizeable contribution to the probability density in the
forward direction along the axis of the incident beam. Calculations carried
out for $^{44}$Ca$(\alpha,\alpha)$ scattering give a very similar picture for
the barrier contribution to the flux. As expected, the internal wave 
contribution to the flux is found to be
nearly negligible in the $^{44}$Ca case, and is not represented here.

\subsection{Divergence of the quantum flux} 
\label{divergence} 

\par The divergence of the flux associated with the scattering wave
function gives a measure of the localization of non-elastic collisions,
which deplete the entrance channel. It is simply related to the probability
density and to the imaginary part $W(r)$ of the optical potential 
used in the calculation by 

\begin{equation}
\label{Difflux}
\ \bbox{\nabla}.\bbox {j}(\bbox {r}) = \frac {2} {\hbar} W(r)
|\psi(\bbox{r})|^2
\end{equation}

which is easily derived from the definition of the flux (Eq. (\ref{flux}))
and from the Schr\"{o}dinger equation.

\par The results obtained for the $\alpha + ^{40}$Ca and $\alpha + ^{44}$Ca
systems are presented in Fig.~\ref{difflux4044}. Inspection of this figure
reveals two contributions to the divergence of the flux.
The first is localized at the outskirts of the potential and
has its maximum on the illuminated side of the target. The second one is
located near
the focus, much deeper inside the potential and is distinctly
much larger in the $\alpha + ^{40}$Ca case. One should of course not forget,
before making any statement about the relative importance of these various
contributions, that integration in three dimensions introduces the factor
$r^2 \sin \theta$ and will have the effect of considerably reducing 
contributions from points located near the origin or near the $z$ axis ($\theta = 0$); in
particular, the contribution of the focus will be much lower than 
Fig.~\ref{difflux4044} would suggest. 

\par In order to disentangle the various contributions to the reaction cross
section, we have calculated the divergence of the quantum flux from the
barrier wave contribution to the scattering wave function, as
also presented in Fig.~\ref{difflux4044}. Since
calculation of the barrier contribution to the wave function involves using
an enhanced absorption, we have taken into account this extra
absorption in the calculation of the
divergence of the barrier flux from Eq. (\ref{Difflux}). As
expected, the barrier contributions for the $\alpha + ^{40}$Ca and $\alpha +
^{44}$Ca systems are found to be very similar and localized at the surface of
the potential. The barrier contribution, $\sigma^{\rm{Reac}}_B$, to the
total reaction cross section can be
obtained by integrating the divergence of the barrier flux over space

\begin{equation}
\label{Reaction}
\ \sigma^{\rm{Reac}}_B=\int d^3r \; \bbox{\nabla.j}_B
\end{equation}

in which unit incident flux has been assumed. More directly, from
the barrier wave $S$-matrix,

\begin{equation}
\label{Reaction2}
\ \sigma^{{\rm{Reac}}}_B=\frac{\pi}{k^2} \sum_\ell
(2\ell+1)(1-|S_{B,\ell}|^2)
\end{equation}

The value obtained for the barrier wave contribution to the $\alpha +
^{44}$Ca reaction cross section (1382 mb) is only 3.5 \% larger than
that
obtained for $\alpha + ^{40}$Ca (1334 mb); this difference is essentially a 
geometrical effect. It is
interesting to calculate in a similar way the internal wave contribution
$\sigma^{{\rm{Reac}}}_I$ to the reaction cross section. This is given in
terms of the internal wave $S$-matrix by     

\begin{equation}
\label{Reaction3}
\ \sigma^{\rm{Reac}}_I=\frac{\pi}{k^2} \sum_\ell (2\ell+1)|S_{I,\ell}|^2
\end{equation}
 
The internal wave contribution to the reaction cross section is found to be
completely negligible in the $^{44}$Ca case (0.04 mb). It has, in contrast, a
modestly larger value in the $^{40}$Ca case (6.8 mb); this last
value represents only about 0.5 \% of the total reaction cross section. It
should not be concluded however that the internal wave does not contribute to
inelastic scattering processes in low energy $\alpha + ^{40}$Ca scattering;
indeed DWBA calculations of the inelastic differential cross section
for excitation of the $J^\pi=3^-, E_x=3.73$ MeV excited state in
$^{40}$Ca, show that use of a strongly absorbing potential for describing the
entrance channel underestimates the inelastic experimental data by more than
an order of magnitude at large angles \cite{Mi77}, and that the spectacular
backward enhancement observed in this inelastic channel is also related to
the internal wave contribution, and thus to the focusing properties of the
potential.

\section{Summary}   
\label{summary}

\par We have calculated the quantum probability density and flux for 
light heavy-ion scattering, taking the $\alpha + ^{40,44}$Ca and $\alpha +
^{16}$O systems as illustrative examples. When absorption is incomplete
($^{40}$Ca and $^{16}$O cases), strong focusing is
observed at low energy, a phenomenon known for a long time in
nucleon-nucleus scattering, and the probability density at the focus is
found to reach
values much higher than that of the incident wave. Classical calculations
then show that the small impact parameter trajectories converge to a point
located near the quantum focus. At low energy these trajectories are
deflected to large angles and the occurence of strong focusing thus appears
to be correlated with the large angle enhancement (ALAS) observed
for these systems. The focus, located well inside the nuclear medium
at low energy, moves away from the illuminated side of the target when the energy
increases and ALAS is progressively replaced by a rainbow behavior.

\par Use of a fully quantal procedure makes possible decomposition of the
scattering wave function into its barrier and
internal wave components, that is, into contributions corresponding
respectively to the part of the incident wave reflected at the
barrier of the effective potential, and to that crossing the barrier and
reemerging after reflection at the innermost turning point. This
decomposition confirms the importance of the focus, which dominates the
internal wave component, in building up the ALAS phenomenon in $\alpha
+ ^{40}$Ca and $\alpha +^{16}$O scattering at low energy. Indeed
for $\alpha + ^{44}$Ca, which is dominated by strong absorption
and where ALAS is absent, the internal wave probability density is found to
be two orders of magnitude lower than that predicted in $\alpha + ^{40}$Ca.
Moreover, the calculation  of the quantum flux for the $\alpha + ^{40}$Ca
system shows that the focusing effect is entirely due to the internal wave
component of the wave function. Finally, calculation of the
divergence of the flux shows that when
absorption is incomplete the focal region gives a sizeable contribution to
non-elastic processes. 

\section*{Acknowledgments}

\par The suggestions of the referee for improvements in our manuscript are gratefully acknowledged.

\begin{figure}
\protect\caption{Comparison of the optical model elastic scattering
differential cross sections (normalized to the Rutherford cross section) for
the $\alpha + ^{40}$Ca (full line) and
$\alpha + ^{44}$Ca (dotted line) systems at 29 MeV incident energy.}
\label{sections4044}
\end{figure}

\begin{figure}
\protect\caption{Probability densities associated with the two optical model
potentials used in Fig.~\ref{sections4044}. In this and similar
figures, the incident beam comes along the negative $z$ axis and the
probability density has been normalized to 1 for large negative $z$ values.}
\label{fo4044}
\end{figure}

\begin{figure}
\protect\caption{Comparison of the $\alpha + ^{40}$Ca and $\alpha + ^{44}$Ca
probability densities at 29 MeV along the axis of the incident beam.}
\label{sagittal4044}
\end{figure}

\begin{figure}
\protect\caption{Quantum probability flux associated with the 29 MeV $\alpha
+ ^{40}$Ca scattering wave function (arbitrary units); in the lower part of
the figure, which presents an enlargement around the focal point, the flux
has been multiplied by a factor of 3. The full-line and dashed-line circles
represent the distances where the real and the imaginary parts of the optical
potential have a depth equal to one tenth of their central values.} 
\label{flux40}
\end{figure}

\begin{figure}
\protect\caption{Same as Fig.~\ref{flux40} for the 29 MeV $\alpha + ^{44}$Ca
system; for that system the real and imaginary radii, as defined in
Fig.~\ref{flux40}, are nearly equal.} 
\label{flux44}
\end{figure}

\begin{figure}
\protect\caption{Classical trajectories for the $\alpha + ^{40}$Ca system at
29 MeV incident energy.}
\label{trajectories}
\end{figure}

\begin{figure}
\protect\caption{Evolution with energy of the probability density for the
$\alpha + ^{16}$O system between 32.2 and 104 MeV.} 
\label{energydep}
\end{figure}

\begin{figure}
\protect\caption{Evolution with energy of the probability density along the
incident beam axis for
the $\alpha+ ^{16}$O system between 32.2 and 146 MeV.}
\label{sagittalo16}
\end{figure}

\begin{figure}
\protect\caption{Modulus of the internal and barrier wave $S$-matrix elements
(upper part) and the corresponding differential cross sections (lower
part) for $\alpha + ^{40,44}$Ca elastic scattering at 29 MeV. (Internal wave
contribution: $^{40}$Ca, dotted line; $^{44}$Ca, dashed line. Barrier
wave
contribution: $^{40}$Ca, full line; $^{44}$Ca, dot-dashed line.)}
\label{decompositions}
\end{figure}

\begin{figure}
\protect\caption{Internal and barrier wave contributions to the probability
density for the $\alpha + ^{40}$Ca system at 29 MeV.}
\label{decompfo40}
\end{figure}

\begin{figure}
\protect\caption{Same as Fig.~\ref{decompfo40} for the $\alpha + ^{44}$Ca
system at 29 MeV.}
\label{decompfo44}
\end{figure}

\begin{figure}
\protect\caption{Internal wave contribution (long dashed lines) and barrier
wave contribution (dotted lines) to the total probability density (full
lines) for the $\alpha + ^{40}$Ca and $\alpha + ^{44}$Ca systems at 29 MeV
along the incident beam axis.} 
\label{sagittalca}
\end{figure}

\begin{figure}
\protect\caption{Internal wave contribution to the probability flux for the
$\alpha + ^{40}$Ca system at 29 MeV. In the upper part, for clarity the
flux in the central region ($r <$ 6 fm) has not been drawn. The
lower part displays the same contribution around the focal point. In the
upper (lower) part of the figure, the flux has been multiplied by a
factor of 1000 (3) with respect to the upper part of Fig.~\ref{flux40}.}
\label{fluxint40}
\end{figure}

\begin{figure}
\protect\caption{Barrier wave contribution to the probability flux for the
$\alpha + ^{40}$Ca system at 29 MeV. The flux has been multiplied by a
factor of 3 with respect to the upper part of Fig.~\ref{flux40}.} 
\label{fluxbar40}
\end{figure}

\begin{figure}
\protect\caption{Divergence of the total (left) and the barrier (right)
probability fluxes for the $\alpha +
^{40}$Ca and $\alpha + ^{44}$Ca systems at 29 MeV in arbitrary units. The barrier
flux has been multiplied by 3 with respect to the total flux.} 
\label{difflux4044}
\end{figure}

\end{document}